\documentclass[12pt]{article}
\usepackage{amsfonts,amsmath,amssymb,epsf,physics,braket,graphicx,hyperref,color,xcolor,comment,bm,tikz}
\usepackage{fancyhdr}
\usepackage{authblk}
\usepackage{mathrsfs}
\usepackage{cite}
\hypersetup{
    unicode=false,          
    pdftoolbar=true,        
    pdfmenubar=true,        
    linktocpage=true,       
    pdffitwindow=false,     
    pdfstartview={FitH},    
    pdfnewwindow=true,      
    colorlinks=true,       
    linkcolor=blue,          
    citecolor=blue,        
    filecolor=blue,      
    urlcolor=blue           
}
\usepackage[top=25truemm,bottom=28truemm,left=20truemm,right=20truemm]{geometry}
\numberwithin{equation}{section}

\title{\bf {\fontsize{20pt}{20pt}\selectfont Bridging dS/CFT and Celestial Holography\\
via Ward-Takahashi Identities}}

\author[1]{
	Hideo~Furugori\thanks{\tt h-furugori(at)gauge.scphys.kyoto-u.ac.jp}
}
\author[2]{
	Naoki~Ogawa\thanks{
	\tt naoki.ogawa(at)yukawa.kyoto-u.ac.jp}
}
\author[1, 3]{
	Sotaro~Sugishita\thanks{
	\tt sotaro(at)gauge.scphys.kyoto-u.ac.jp}
	\vspace{5mm}
}
\author[2]{
	Takahiro~Waki\thanks{
	\tt takahiro.waki(at)yukawa.kyoto-u.ac.jp}
}
\affil[1]{\it\normalsize Department of Physics, Kyoto University, Kyoto 606-8502, Japan}
\affil[2]{\it\normalsize Center for Gravitational Physics and Quantum Information,\protect\\
Yukawa Institute for Theoretical Physics, Kyoto University, \protect\\
Kitashirakawa Oiwakecho, Sakyo-ku, Kyoto 606-8502, Japan}
\affil[3]{\it\normalsize RIKEN iTHEMS, Wako, Saitama 351-0198}
\date{}

\usetikzlibrary{calc}

\newcommand{\be}{\begin{equation}}           
\newcommand{\ee}{\end{equation}}             
\newcommand{\nn}{\nonumber \\}  

\numberwithin{equation}{section}  

\usepackage{amsthm}

\begin{document}
\maketitle
\thispagestyle{fancy}
\renewcommand{\headrulewidth}{0pt}

\begin{abstract}
In \cite{Furugori:2025xkl}, we provide a map from a scalar theory on $(D+2)$-dimensional Minkowski spacetime to a scalar theory with a continuous mass spectrum on $(D+1)$-dimensional de Sitter spacetime, and propose a link between celestial amplitudes and cosmological correlators (the cosmological–celestial dictionary).
We extend the construction to fields with spin 1 and 2, and find that massless spin fields map to spin fields with continuous mass spectra. 
In this construction, we identify the de Sitter counterparts of the Nambu-Goldstone modes associated with the asymptotic symmetries in Minkowski spacetime. 
For $U(1)$ gauge theories, the counterpart is restricted to the massless sector within the continuous Proca spectrum, while for linearized gravity supertranslations are encoded in the partially massless sector and superrotations in the strictly massless sector. 
Using the identification, we reveal that the associated Ward--Takahashi identities of the cosmological correlators reproduce the conformally soft photon and graviton theorems via the cosmological–celestial dictionary. 
In particular, the celestial stress tensor is derived from the asymptotic limit of gravitons in de Sitter spacetime.
\end{abstract}

\newpage
\thispagestyle{empty}
\setcounter{tocdepth}{2}

\setlength{\abovedisplayskip}{12pt}
\setlength{\belowdisplayskip}{12pt}

\tableofcontents
\newpage

\section{Introduction}
The holographic principle \cite{tHooft:1993dmi, Susskind:1994vu} has seen remarkable development in recent years and now provides a powerful guide to understanding quantum gravity. Its most thoroughly explored realization is the AdS/CFT correspondence \cite{Maldacena:1997re, Gubser:1998bc, Witten:1998qj}, whose validity has been tested across a wide range of setups and explicit computations. In its standard form, AdS/CFT posits an equivalence between a quantum gravity theory on $(d+1)$-dimensional anti-de Sitter spacetime and a $d$-dimensional conformal field theory.
Motivated by this success, attention has turned to holographic dualities beyond AdS. In particular, the dS/CFT correspondence has been actively investigated as a putative duality between quantum gravity in $(d+1)$-dimensional de Sitter spacetime and a $d$-dimensional CFT living on the conformal boundary \cite{Witten:2001kn, Strominger:2001pn, Maldacena:2002vr}. 
Parallel to these efforts, intensive work has also focused on holography for asymptotically flat spacetimes, often referred to as celestial holography \cite{Kapec:2016jld, Pasterski:2016qvg, Strominger:2017zoo, Pasterski:2017kqt, Pasterski:2017ylz, Donnay:2020guq, Ogawa:2022fhy}. This program proposes to recast scattering amplitudes in $(d+1)$-dimensional flat space as correlation functions on the celestial sphere $S^{d-1}$, and concrete examples have been worked out in a variety of theories.

\begin{figure}
    \centering
    \includegraphics[width=0.8\linewidth]{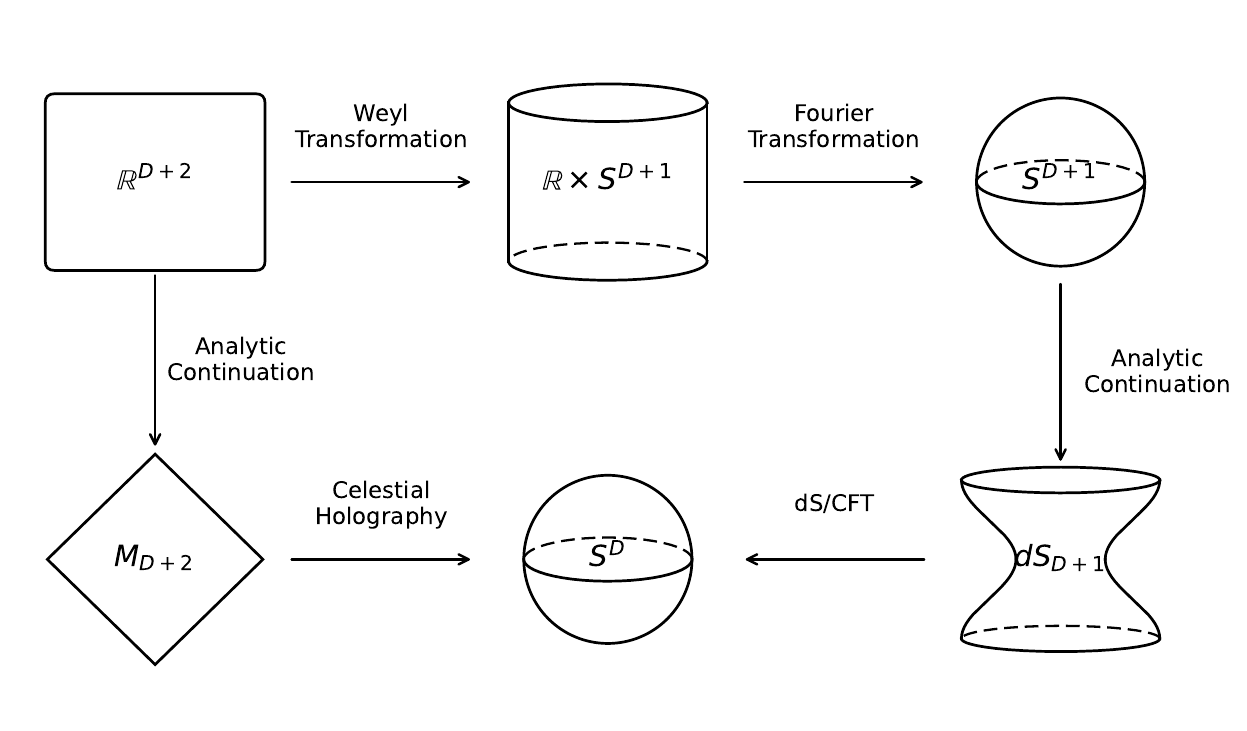}
    \caption{A schematic diagram illustrating how celestial holography meets the dS/CFT correspondence \cite{Furugori:2025xkl}. }
    \label{fig1}
\end{figure}

Celestial holography still contains many unexplored aspects, and various attempts have been made to interpret it through the lens of other holographic frameworks \cite{Casali:2022fro, Sleight:2023ojm, Ogawa:2022fhy, Ogawa:2024nhx, Hao:2025ocu}. In our previous work \cite{Furugori:2025xkl}, we proposed an interpretation of celestial holography from the perspective of the dS/CFT correspondence. Starting from a massless scalar field theory on flat space $\mathbb{R}^{D+2}$, we performed a Weyl transformation to recast the geometry as $S^{D+1}\times \mathbb{R}$. By subsequently applying a Fourier transformation along the $\mathbb{R}$ direction, the theory can be rewritten as one defined on $S^{D+1}$ with a continuous mass spectrum. Upon analytic continuation, this construction establishes an equivalence between the massless scalar theory on Minkowski space $\mathbb{M}_{D+2}$ and a theory on de Sitter space $\mathrm{dS}_{D+1}$ possessing a continuous mass spectrum with the Bunch--Davies vacuum \cite{Bunch:1978yq, Mottola:1984ar, Allen:1985ux, Marolf:2010zp, Marolf:2010nz, Higuchi:2010xt} (see Fig.~\ref{fig1}). Through this correspondence, one can identify operators in the celestial conformal field theory (CCFT) with asymptotic cosmological operators in de Sitter space. Compared to the framework that partitions $\mathbb{M}_{D+2}$ into two Euclidean AdS regions and one dS region \cite{Cheung:2016iub, deBoer:2003vf, Ball:2019atb,  Iacobacci:2022yjo, Sleight:2023ojm, Iacobacci:2024nhw, Jorstad:2023ajr}, our approach offers a clearer physical interpretation and a more natural foundation for developing celestial holography as a genuine holographic duality.

In our previous work \cite{Furugori:2025xkl}, we built the correspondence (cosmological-celestial dictionary) between celestial scalar operators $\mathscr{O}_{\Delta}$ and cosmological operators $V_{\Delta}, W_{\Delta}$ without spin.
More concretely, we consider the late/early time limit of the $\mathrm{dS}$ bulk scalar field  $\psi_\lambda$ that behaves as 
\begin{align}
\label{naive:asympt_behavior_dS_field}
\psi_\lambda \sim e^{\mp(D-\Delta) t}V_{D-\Delta}+e^{\mp\Delta t}W_{\Delta}
\qquad (t\rightarrow \pm\infty).
\end{align}
Here, $ \Delta={D}/{2}+ i\lambda$.
We define the operators $V_\Delta$ and $W_\Delta$ as the coefficients of these asymptotic behaviors. 
In this paper, we refer to these extrapolated operators as cosmological operators.
In addition, we define the  celestial operator $\mathscr{O}_{\Delta}$ by the Mellin transformation from the annihilation/creation operators of the Minkowski scalar field as follows \cite{Pasterski:2016qvg}:
\begin{align}
\mathscr{O}_\Delta(p)=\int_0^\infty \frac{d\omega}{\omega} \omega^{\Delta} a(\omega p).
\end{align}
These $V_{\Delta},W_{\Delta}$ can be regarded as operators on ${S}^{D}$. We showed the relation between the two operators as
\begin{align}
\begin{aligned}
V_{\Delta}(p)
&= N_{\lambda}i^{-\Delta}\widetilde{\mathscr{O}_{D-\Delta}}(p),\\
W_{\Delta}(p)
&= N_\lambda i^{-\Delta}\mathscr{O}_{\Delta}(p), 
\end{aligned}
\label{dS_celestial_scalar}
\end{align}
where $N_\lambda = \frac{\Gamma(-i\lambda)}{2^{5/2}\pi^{D/2+3/2}}$. Now we slightly change the definition of cosmological operators and celestial operators from our previous work \cite{Furugori:2025xkl} in order to obtain a simple relation as \eqref{dS_celestial_scalar}. 
A precise definition can be seen in \eqref{asympt_behavior_dS} and \eqref{celestial_scr}.

In this paper, we extend the above analysis to gauge fields on $\mathbb{M}_{D+2}$. 
We find that after performing the same sequence of transformations, a $U(1)$ gauge theory on $\mathbb{M}_{D+2}$ is mapped to a Proca field on $\mathrm{dS}_{D+1}$ with a continuous mass spectrum, while linearized gravity becomes a Fierz--Pauli field with a continuous mass spectrum. 
An important question is how the gauge degrees of freedom in $\mathbb{M}_{D+2}$ manifest themselves in the $\mathrm{dS}_{D+1}$ description. 
For the $U(1)$ case, we show that the gauge symmetry corresponds precisely to that for a spin-1 gauge field on de Sitter space. 
For linearized gravity, the asymptotic symmetries in $\mathbb{M}_{D+2}$ consist of supertranslation and superrotation modes. 
We demonstrate that they correspond respectively to symmetries for the partially massless and massless gravitons on $\mathrm{dS}_{D+1}$. 
In celestial holography, the Ward-Takahashi identities associated with supertranslation and superrotation are expressed in terms of conformally soft operators with conformal dimensions $\Delta = 1$ and $\Delta = 0$. 
On $\mathrm{dS}_{D+1}$, the partially massless Fierz--Pauli field and the massless spin--2 field correspond to operators with $\Delta = 1$ and $\Delta = 0$, respectively, and we find that the Ward-Takahashi identities associated with the symmetry coincide with those of celestial holography.

This paper is organized as follows.
In Sec.~\ref{Sec.3}, we perform a kind of ``dimensional reduction'' of gauge fields in flat spacetime to fields on de~Sitter space, both for the Maxwell theory and linearized gravity. By examining how the original gauge transformations are realized after this mapping, we show that they act on the massless and partially massless sectors on de~Sitter space.
In Sec.~\ref{Sec.4}, we construct a correspondence between operators in Minkowski space and cosmological operators on de~Sitter space by extending the analysis of~\cite{Furugori:2025xkl} to fields with spin.
In Sec.~\ref{Sec.5}, we study the Ward-Takahashi identities of the cosmological correlators and demonstrate that, via the correspondence established in Sec.~\ref{Sec.4}, they reproduce the Ward-Takahashi identities of celestial holography, which are equivalent to soft theorems in flat spacetime.
Finally, Sec.~\ref{Sec.6} is devoted to a summary and discussions.

\subsection{Useful formula and some notation}
In this section, we show some useful formulas and notations used in this paper.

\paragraph{Embedding formalism}
We define the measure $D^D p$ of conformal integrals followed by \cite{Simmons-Duffin:2012juh} 
\begin{align}
    \int D^Dp~ f(p)
    &\equiv \frac{2}{\mathrm{Vol}\,\mathrm{GL}(1,\mathbb{R})^+}
      \int_{p^0>0} d^{D+2}p \,\delta\left(p^2\right)f(p),
\end{align}
where $f(p)$ should have a degree $-D$ under the rescaling $p \to a p$.
If we take a ``gauge'' $p=(1,\bm{p})$ where $\bm{p}$ is a unit vector in $\mathbb{R}^{D+1}$, we have  
\begin{align}
    \int D^D p\,f(p)
    =\int_{\mathrm{S}^D} d^D\bm{p} f(p)\Big| _{p=(1,\bm{p})}.
\end{align}
The shadow transformation for scalar operator $\widetilde{\mathcal{O}_\Delta}$ is defined by 
\begin{align}\label{def:shadow}
    \widetilde{\mathcal{O}_{\Delta}}(p)
  = \frac{1}{S_{\Delta}}
    \int d^D k (-2p \cdot k)^{\Delta-D}
    \mathcal{O}_{\Delta}(k).
\end{align}
The normalization factor $S_{\Delta} $ is given by
\begin{align}
  S_{\Delta}
  = 2^{2\Delta-D}\pi^{h}
    \frac{\Gamma(\Delta-h)}{\Gamma(D-\Delta)}, 
  \qquad
  h = \frac{D}{2},
\end{align}
which is chosen such that the transform is involutive\footnote{The involutive condition \eqref{involutive} does not uniquely fix $S_{\Delta}$. For instance, we may choose  $S_{\Delta} = \pi^{h}
    \frac{\Gamma(\Delta-h)}{\Gamma(D-\Delta)}$. We add the factor $2^{2\Delta-D}$ for later convenience.}:
\begin{align}
  \label{involutive}
  \widetilde{\left(\widetilde{\mathcal{O}_{\Delta}}\right)}
  = \mathcal{O}_{\Delta}.
\end{align}
The definition of the shadow transformation can be extended for spinning particles as follows:
\begin{align}
    \widetilde{\mathcal{O}_{I_1,\dots I_l, \Delta}}(p)
  = \frac{1}{S_{\Delta}}
    \int d^D k (-2p \cdot k)^{\Delta-D}
    \prod_{n=1}^l \text{Pr}^{J_n}{}_{I_n}(p,k)\mathcal{O}_{J_1,\dots J_l, \Delta}(k),
\end{align}
where $\text{Pr}$ is a projection tensor defined by
\begin{align}
\text{Pr}^J{}_I(X,Y)= \delta^J{}_I 
- \frac{(Y \cdot Y)\,X^J X_I
-(X \cdot Y)\,\bigl(X^J Y_I+Y^J X_I\bigr)
+(X \cdot X)\,Y^J Y_I}{ (X \cdot X)(Y \cdot Y)-(X \cdot Y)^2}.
\end{align}

\paragraph{Coordinate and metric}
We begin by considering the $(D{+}2)$-dimensional Euclidean space $\mathbb{R}^{D+2}$ with Cartesian coordinates $X^I$, where indices $I,J,\dots$ are used and raised/lowered with the flat metric $G_{IJ}=\delta_{IJ}$.

Next, we introduce intrinsic coordinates $x^i$ on the unit sphere $S^{D+1}$, with indices $i,j,\dots$ raised and lowered by the sphere metric $\gamma_{ij}$. We also use embedding coordinates $x^I$ for $S^{D+1}$ subject to the constraint $x^Ix_I=1$.

The fundamental metric is the flat metric on $\mathbb{R}^{D+2}$, given by
\begin{align}
ds^2 = G_{IJ}\,dX^I dX^J .
\label{Cartesian}
\end{align}
Introducing polar coordinates $X^I = r\,x^I$ (with $r>0$, $x\in S^{D+1}$), we obtain
\begin{align}
ds^2 = dr^2 + r^2 \gamma_{ij}\,dx^i dx^j 
      = e^{2\xi}\big(d\xi^2 + \gamma_{ij}\,dx^i dx^j\big),
\qquad r=e^{\xi}.
\label{Polar}
\end{align}
Thus, by a Weyl transformation, the flat metric can be mapped to the cylindrical form as
\begin{align}
e^{-2\xi}G_{IJ}\,dX^I dX^J = d\xi^2 + \gamma_{ij}\,dx^i dx^j .
\end{align}

In addition as in \cite{Furugori:2025xkl}, we parametrize the coordinates on $\mathrm{M}_{D+2}$ as
\begin{align}
X(t, r, \bm{p}) &= \left(r\sinh t, r\cosh t \,\bm{p}\right),\label{param_Mink}
\end{align}
where $\bm{p}$ is a unit vector, which can be identified with a point on $\mathrm{S}^{D}$.  
We parametrize the unit sphere $\mathrm{S}^{D}$ by coordinates $x^\alpha$ and denote its metric by $\gamma_{\alpha\beta}$.

Similarly we parametrize dS$_{D+1}$ coordinates as follows: 
\begin{align}
x(t, r, \bm{p}) &= \left(\sinh t, \cosh t \bm{p}\right). 
\label{param_dS}
\end{align}
For convenience, we decompose the vector $X$ into null vectors $p_\pm = (1,\pm\bm{p})$ as
\begin{align}
X = \frac{1}{2}r e^{t}p_+  - \frac{1}{2}r e^{-t}p_-, 
\end{align}
and similarly we define dS$_{D+1}$ embedding coordinates $x$ decompose as follows: 
\begin{align}
x = \frac{1}{2} e^{t}p_+  - \frac{1}{2} e^{-t}p_-.
\label{de_sitter_paramet}
\end{align}

\paragraph{Definition of cosmological operators}
Then, the extrapolated operators in dS are defined as the leading coefficients of the dS bulk field in the late/early time limit:
\begin{align}
\label{asympt_behavior_dS}
\begin{aligned}
\psi_\lambda(x) &\sim e^{-(D-\Delta) t}V_{D-\Delta}(p)+e^{-\Delta t}W_{\Delta}( p) \quad\quad\quad~~~ (t\to \infty)\\
\psi_\lambda(x) &\sim e^{-(D-\Delta) t}V_{D-\Delta}(-p_-)+e^{-\Delta t}W_{\Delta}(-p_-)\quad (t\to-\infty)
\end{aligned}
\end{align}
where we have used the parametrization \eqref{de_sitter_paramet} de Sitter coordinates  with $p=p_+$.

In our previous work~\cite{Furugori:2025xkl}, the convention $\lambda > 0$ was imposed. In this paper, we instead allow $\lambda \in \mathbb{R}$. With this convention, we should use the symbols $V_\Delta$ and $W_{\Delta}$ (instead of $O_{\Delta_+}$ and $O_{\Delta_-}$ used in our previous paper) to make explicit that $O_{\Delta_+}$ and $O_{\Delta_-}$ are not necessarily related to each other by analytic continuation. The relation to the conventions used in our previous work~\cite{Furugori:2025xkl} is as follows:
\begin{align}
&V_{\Delta_- } \equiv O_{\Delta_-}, \qquad
W_{\Delta_+} \equiv O_{\Delta_+},\
&V_{\Delta_+ } \equiv \bigl(O_{\Delta_-}\bigr)^\dagger, \qquad
W_{\Delta_- } \equiv \bigl(O_{\Delta_+}\bigr)^\dagger.
\end{align}
\paragraph{Definition of celestial operators} We use the following notation to avoid notational complexity due to the in/out index and make the analyticity of the complex-$p$ plane manifest, namely, crossing symmetry.
First, we define momentum operators as follows: 
\begin{align}
\begin{aligned}
a(\omega p)&\equiv a_{\text{out}}(\omega \bm{p}) \\
a(-\omega p)&\equiv a^\dagger_{\text{in}}(\omega \bm{p}).
\end{aligned}
\end{align}

We also define the celestial operators as follows by using the above momentum operator.\footnote{
Note that we now allow the embedding coordinate $p$ of the operator $\mathscr{O}$ to be past-directed. Therefore, this is not the standard definition of an embedding-space CFT operator. Although this is merely a matter of convention, we can recover the usual CFT operator by  \eqref{celestial_dic}.
}
\begin{align}
\mathscr{O}_\Delta(p)=\int_0^\infty \frac{d\omega}{\omega} \omega^{\Delta} a(\omega p)
\label{celestial_scr}
\end{align}

Now we would like to connect the convention of our previous work~\cite{Furugori:2025xkl}. In the celestial holography, we adopt the rectified dictionary \cite{Furugori:2023hgv}. It is supposed that the creation and annihilation operators of bulk fields on the flat space are related to the primary and their shadow operators in the celestial CFT as
\begin{align}\label{celestial_dic}
\begin{aligned}
  \mathcal{O}_{\Delta}^{+}(p)
    &= \mathscr{O}_\Delta(p)\\
  \widetilde{\mathcal{O}_{D-\Delta}^{-}}(p)
    &= \mathscr{O}_\Delta(-p),
\end{aligned}
\end{align}
where we use simple notation.

\paragraph{Generalized Delta Function}
Following \cite{Donnay:2020guq}, we introduce the generalized delta function as an extension of the usual delta function.  
Let $f(x)$ be a test function on $\mathbb{C}$, and for a complex number $z \in \mathbb{C}$ define
\begin{align}
    f(z)
    \;=\; \int_{-\infty}^{\infty} dx\; f(x)\,\delta^{G}(x - z).
\label{eq:def_gene}
\end{align}
A distribution $\delta^{G}$ that satisfies this relation will be called the generalized delta function.

When the test function has the analyticity properties that will be specified shortly, the generalized delta function $\delta^{G}(z)$ admits the following useful representation as a distribution:
\begin{align}
    \delta^{G}(z)
    \;=\; \frac{1}{2\pi}\int_{-\infty}^{\infty} dt\; e^{i t z}.
\label{eq:exp:generalized_delta}
\end{align}

One can verify that this expression indeed satisfies the definition~\eqref{eq:def_gene} as follows.
Writing $z = x' + i y'$ with $x', y' \in \mathbb{R}$, we obtain
\begin{align}
  &\frac{1}{2\pi}\int_{-\infty}^{\infty} dx\int_{-\infty}^{\infty} dt\;
       f(x)\,e^{i t (x - (x' + i y'))}
       \nonumber\\
    =& \frac{1}{2\pi}\int_{-\infty + i y'}^{\infty + i y'} dx\int_{-\infty}^{\infty} dt\;
       f(x)\,e^{i t (x - (x' + i y'))}
       \nonumber\\
    =& \frac{1}{2\pi}\int_{-\infty}^{\infty} dx\int_{-\infty}^{\infty} dt\;
       f(x + i y')\,e^{i t (x - x')}
       \nonumber\\
    =& \int_{-\infty}^{\infty} dx\; f(x + i y')\,\delta(x - x')\\
    =& f(z).
\end{align}
In going from the first to the second line, we have deformed the contour of the $x$-integration from the real axis to the line $\Im x = y'$.
This deformation is justified provided that $f$ is analytic in the strip $0 \leq \Im x \leq y'$ and decays sufficiently fast at infinity.
Therefore, when acting on such test functions, the generalized delta function can be represented in the form~\eqref{eq:exp:generalized_delta}.

\section{Dimensional Reduction of Maxwell Theory and Gravity}\label{Sec.3}
In this section, following \cite{Furugori:2025xkl,Cheung:2016iub}, we perform a radial decomposition of the free Maxwell theory and, in parallel, the linearized gravity in $(D{+}2)$-dimensions.
We also consider transformations in the $(D{+}1)$-dimensional theory associated with the gauge transformation in $(D{+}2)$-dimensions.

\subsection{Maxwell}
We start from the free $U(1)$ gauge theory on $\mathbb{R}^{D+2}$,
\begin{align}
  S  =  \frac{1}{4}\int_{\mathbb{R}^{D+2}} [d^{D+2}X] F_{IJ}F^{IJ}.
  \label{maxwell}
\end{align}
where $F_{IJ}:=\partial_IA_J-\partial_JA_I$. To obtain a canonically normalized theory on the cylinder $S^{D+1}\times\mathbb{R}_\xi$, it is convenient to work with the Weyl-rescaled field
\begin{align}
\mathcal{A}_I= e^{\frac{D-2}{2}\xi} A_I,
\label{emweyl}
\end{align}
and imposed the radial gauge condition $\mathcal{A}_\xi=0$.
After the Weyl transformation, the action becomes
\begin{align}
  S  =  \frac{1}{4}\int_{\mathbb{R}} d\xi \int_{S^{D+1}}[d^{D+1}x]\,
  \qty(\mathcal{F}_{ij}\mathcal{F}^{ij}+2\qty(\partial_\xi \mathcal{A}-\frac{D-2}{2}\mathcal{A})^2),
  \label{eq:maxwell-cylinder-D}
\end{align}
where indices are raised with $\gamma^{ij}$ and
$\mathcal{F}_{ij}:=\nabla_i\mathcal{A}_j-\nabla_j\mathcal{A}_i$. 

Next, we perform the Fourier transformation along $\xi$-direction:
\begin{align}
  a_i(\lambda,x)  =  \frac{1}{\sqrt{2\pi}}\int_{-\infty}^{\infty}  d\xi  e^{-i\lambda \xi}\,\mathcal{A}_i(\xi,x), 
  \qquad
  \mathcal{A}_i(\xi,x)  =  \frac{1}{\sqrt{2\pi}}\int_{-\infty}^{\infty}  d\lambda  e^{+i\lambda \xi}\,a_i(\lambda,x),
\end{align}
with the reality condition $a_i(-\lambda,x)=a_i(\lambda,x)^\dagger$.  
We then yield a continuum of complex vector fields on $S^{D+1}$ labeled by $\lambda$:
\begin{align}
  S  =  \int_{S^{D+1}}[d^{D+1}x]\int_{-\infty}^{\infty} d\lambda 
  \left[\frac{1}{4}\,f_{ij}(\lambda)f^{ij}(-\lambda) 
  + \frac{1}{2}\,\Big(\lambda^2+\frac{(D-2)^2}{4}\Big)\,a_i(\lambda)a^i(-\lambda)\right],
  \label{eq:vector-lambda-action-D}
\end{align}
where $f_{ij}(\lambda):=\nabla_i a_j(\lambda)-\nabla_j a_i(\lambda)$.
Each mode thus obeys a Proca equation on $S^{D+1}$ with Kaluza–Klein mass
\begin{align}
  m_\lambda^2=\lambda^2+\frac{(D-2)^2}{4}.
\end{align}
This organizes the modes into principal series representations with conformal dimensions of the boundary correlator,
\begin{align}
  \Delta_{\pm}=\frac{D}{2}\pm i\lambda,
\end{align}
which is consistent with the relation of conformal dimension for massive spin-$\ell$ particle \cite{Giombi:2009wh} for symmetric traceless tensors,
\begin{align}
-m^2=(\Delta+\ell-2)(\Delta-\ell-D+2)
\label{casi}
\end{align}
evaluated at $\ell=1$.
This realizes the dimensional reduction of the free spin-1 field on $\mathbb{R}^{D+2}$ 
as a tower of vector modes on $S^{D+1}$ with a continuous mass spectrum.

Finally we note the relationship between $A$ and $a$. 
\begin{align}
  a_i(\lambda,x) 
  & =  \frac{1}{\sqrt{2\pi}}\int_{-\infty}^{\infty}  d\xi  e^{\frac{D-2}{2}\xi}e^{-i\lambda\xi} \,A_i\big(X(\xi,x)\big)\nn
  & =  \frac{1}{\sqrt{2\pi}}\int_{0}^{\infty} \frac{dr}{r} r^{\Delta_- -1}  \,A_i(X(r,x)).
  \label{maxmiyamiya}
\end{align}
For convenience, we sometimes use the embedding formalism. 
We uplift the index of $a_i$ as $(D+2)$-dimensional index $a_I$ as follows:
\begin{align}
   \frac{\partial x^I}{\partial x^i} a_I\equiv a_i ,
   \label{uplifting_maxwell}
\end{align}
and we impose $x^I a_I=0$.
We then have
\begin{align}
   a_I(\lambda,x) 
& =  \frac{1}{\sqrt{2\pi}}\int_{0}^{\infty} \frac{dr}{r} r^{\Delta_- }A_I(X(r,x))
\label{maxmiya}
\end{align}
where we use definition $A_i=\frac{\partial X^I}{\partial x^i} A_I=r\frac{\partial x^I}{\partial x^i} A_I$.

\subsubsection{Gauge symmetry in the sphere action}
Let us consider a symmetry of the action \eqref{eq:vector-lambda-action-D}. 
The original theory has a gauge symmetry $\delta A_I = \partial_I \Lambda$ where $\Lambda$ is independent of $\xi$ because we impose the radial gauge $A_\xi=0$. 
For the rescaled field $\mathcal{A}$, the transformation corresponds to
\begin{align}
  \delta \mathcal{A}_i(\xi,x)=\partial_i\Lambda(x)\,e^{\frac{D-2}{2}\xi}.
\end{align}
However, the action \eqref{eq:maxwell-cylinder-D} is invariant up to boundary terms under the replacement
$\left(\partial_\xi \mathcal{A} - \frac{D-2}{2}\mathcal{A}\right)^2
\to \left(\partial_\xi \mathcal{A} + \frac{D-2}{2}\mathcal{A}\right)^2,$
and so the action is also invariant under the transformation\footnote{
This transformation is a symmetry of the action on the sphere, but it does not constitute a gauge symmetry of the original flat-space action. Nevertheless, it maps a given configuration to a distinct solution with nonvanishing field strength, and therefore represents a genuine symmetry.
In this sense, this mode is not a pure gauge mode in the original action. 
However, since this corresponds to a pure gauge mode in the de Sitter action, we shall also refer to it as a “pure gauge mode” throughout this paper.
}
\begin{align}
  \delta \mathcal{A}_i(\xi,x)=\partial_i\Lambda(x)\,e^{-\frac{D-2}{2}\xi}.
\end{align}
This property essentially originates from the inversion symmetry
$X^{I} \to X^{I}/|X|^{2}$, namely $\xi\to -\xi$,
present in \eqref{maxwell}.\footnote{
It is important to note that this inversion symmetry becomes exact only in $D=2$.
For generic dimensions $D \neq 2$, the inversion transformation yields an overall Weyl factor, but this is compensated for by the Weyl transformation specified in \eqref{emweyl}. $D=2$ case is discussed by \cite{Jorstad:2023ajr}.
}
Therefore, the action of \eqref{eq:vector-lambda-action-D} has gauge symmetry as
\begin{align}
  \delta_\Lambda a_i(\lambda,x)
  = \sqrt{2\pi} \delta^{G} \qty(\lambda\pm\frac{i}{2}(D-2)) \partial_i\Lambda(x).
  \label{eq:pure-gauge-delta}
\end{align}
Thus, pure-gauge modes localize at $\lambda=\pm i(D-2)/2$, i.e., the ``massless modes''.
We should note that for $D\neq 2$, the massless modes $\lambda=\pm i(D-2)/2 $ are absent in the spectrum \eqref{eq:vector-lambda-action-D}.
However, by performing an analytic continuation, we can treat these modes.  
In the context of celestial holography, this corresponds to defining the conformally soft mode by analytically continuing the conformal wave functions~\cite{Donnay:2018neh}.
Indeed, in Sec.~\ref{Sec.5} we will see that the Ward-Takahashi identity associated with these modes is equivalent to the conformally soft theorem.

In later sections, we can see that the choice of sign is equivalent to deciding whether the cosmological operators $V_{\Delta_-}$ and $W_{\Delta_+}$, associated with a massless particle, are assigned conformal dimensions $D-1$ or $1$, respectively.\footnote{In the case 
$D=2$, these two modes become degenerate.}
In other words, this is equivalent to choosing which of the two modes is identified as the celestial operator and which as its shadow operator.
We finally note that the relation between inversion symmetry in Minkowski spacetime and the shadow transformation in CFT is discussed in detail in \cite{Jorstad:2023ajr}.

\subsection{Gravity}\label{subsec:2gravity}
We now turn to the gravitational sector. Our starting point is the $(D{+}2)$-dimensional Einstein-Hilbert action with the vanishing cosmological constant
\begin{align}
    S  = \frac{1}{2}\int_G [d^{D+2}X]\,R'[G]\,,
    \label{eq:Weyl-inv-parent}
\end{align}
where $R'[G]$ is the Ricci scalar of the $(D{+}2)$-dimensional metric $G_{IJ}$.  
We work in the radial gauge and parametrize the metric as
\begin{align}
  G_{IJ}dX^I dX^J
   =  e^{2\xi}\bigl(d\xi^2+g_{ij}(\xi,x)\,dx^i dx^j\bigr).
  \label{eq:radial-metric}
\end{align}
We now perform the Weyl transformation
\begin{align}
    g_{IJ}  =  e^{-2\xi}G_{IJ},
    \label{Weylmetric}
\end{align}
so that $g_{IJ}dX^I dX^J = d\xi^2 + g_{ij}(\xi,x)\,dx^i dx^j$.  
Using the standard ADM decomposition of the scalar curvature, the action \eqref{eq:Weyl-inv-parent} becomes
\begin{align}
    S 
    &= \frac{1}{2}\int d\xi\, e^{D\xi} \int_g [d^{D+1}x]\,
    \Bigl(R[g] - D(D-1) + K^2 - K_{ij}K^{ij}\Bigr),
    \label{eq:EH-on-cylinder}
\end{align}
where $R[g]$ is the Ricci scalar of the $(D{+}1)$-dimensional metric $g_{ij}$ and $K_{ij}$ is the extrinsic curvature of the constant-$\xi$ slices,
\begin{align}
  K_{ij}  =   \frac12\,\partial_\xi g_{ij},\qquad
  K  =  g^{ij}K_{ij}.
\end{align}
The detailed derivation of \eqref{eq:EH-on-cylinder} is given in App.~\ref{App.gravreduction}.  
Thus each $\xi$-slice is governed by Einstein–Hilbert dynamics in $(D{+}1)$-dimensions with a positive cosmological constant $\Lambda=\frac12 D(D-1)$.

We now extract the mass spectrum of this theory.  
Let $g_{ij}=\gamma_{ij}+H_{ij}$ and $H=\gamma^{ij}H_{ij}$.  
Expanding \eqref{eq:EH-on-cylinder} to quadratic order in $H_{ij}$, we obtain
\begin{align}
S^{(2)}
&= \frac{1}{8}\int d\xi\, e^{D\xi}\int_{S^{D+1}} [d^{D+1}x]\,
\Bigl(
L_{FP}(H)
- \qty(  \partial_\xi H_{ij}\,\partial_\xi H^{ij}-\qty(\partial_\xi H)^2)
\Bigr),
\label{eq:quad-action-H}
\end{align}
where all covariant operations are taken with respect to $\gamma_{ij}$, indices are raised with $\gamma^{ij}$, and
\begin{align}
\begin{aligned}
    L_{FP}(H)
    &= -\nabla_k H_{ij}\,\nabla^k H^{ij}
    + 2\nabla_i H_{jk}\,\nabla^j H^{ik}
    - 2\nabla_i H^{ij}\,\nabla_j H \\
    &\quad\quad\quad
    + \nabla_k H\,\nabla^k H
    +2D\qty(H_{ij}H^{ij}-\frac12 H^2)
\end{aligned}
\end{align}
is the Fierz–Pauli Lagrangian on $S^{D+1}$ (see App. \ref{FPreview}). The last term is not a mass term; rather, it originates from the fact that the background spacetime is a sphere \cite{Aragone:1971kh, Aragone:1979bm, Kan:2022fcj}.

We rescale the fields as follows to absorb the overall factor $e^{D\xi}$,
\begin{align}
    \mathcal{H}_{ij} = e^{\frac{D}{2}\xi} H_{ij},
\end{align}
and obtain
\begin{align}
S^{(2)}
&= \frac{1}{8}\int d\xi \int_{S^{D+1}} [d^{D+1}x]\,
\Bigl(
L_{FP}(\mathcal{H})
- \bigl(\partial_\xi \mathcal{H}_{ij}\,\partial_\xi \mathcal{H}^{ij}-(\partial_\xi \mathcal{H})^2\bigr)
- \frac{D^2}{4}\bigl(\mathcal{H}_{ij}\mathcal{H}^{ij}-\mathcal{H}^2\bigr)
\Bigr),
\end{align}
where $\mathcal{H}=\gamma^{ij}\mathcal{H}_{ij}$.

Next, we perform a Fourier transform along the $\xi$-direction,
\begin{align}
\begin{aligned}
    h_{ij}(\lambda,x)
    &= \frac{1}{\sqrt{2\pi}}\int_{-\infty}^{+\infty} d\xi \,e^{-i\lambda \xi}\, \mathcal{H}_{ij}(\xi,x),\\
    \mathcal{H}_{ij}(\xi,x)
    &= \frac{1}{\sqrt{2\pi}}\int_{-\infty}^{+\infty} d\lambda \,e^{+i\lambda \xi}\, h_{ij}(\lambda,x),
\end{aligned}
\end{align} 
with the reality condition $h_{ij}(-\lambda,x)=h_{ij}(\lambda,x)^\dagger$.  
Inserting this into the action, we obtain
\begin{align}
S^{(2)}
&= \frac{1}{8}\int_{-\infty}^\infty d\lambda \int_{S^{D+1}} [d^{D+1}x]\,
\Bigl(
L_{FP}^C(h(\lambda))
- m^2_\lambda\bigl(h_{ij}(\lambda)h^{\dagger ij}(\lambda)-h(\lambda)h^\dagger(\lambda)\bigr)
\Bigr),
\end{align}
where
\begin{align}
    m_\lambda = \sqrt{\frac{D^2}{4}+\lambda^2},
\end{align}
and
\begin{align}
\begin{aligned}
    L_{FP}^C(h)
    &= -\nabla_k h_{ij}^\dagger\,\nabla^k h^{ij}
    + 2\nabla_i h^{\dagger }_{jk}\,\nabla^j h^{ik}
    - 2\nabla_i h^{\dagger ij}\,\nabla_j h \\
    &\quad
    + \nabla_k h^\dagger\,\nabla^k h
    + 2D\qty(h_{ij}^\dagger h^{ij}-\frac12 h^\dagger h),
\end{aligned}
\end{align}
with $h = \gamma^{ij}h_{ij}$.  

As for the vector modes, using \eqref{casi}, we can read off the conformal dimensions associated with these modes as
\begin{align}
    \Delta_\pm = \frac{D}{2} \pm i\lambda.
\end{align}

Finally, it is often convenient to trade the coordinate $\xi$ for a radial variable $r = e^{\xi}$ and rewrite the Fourier transform as a Mellin transform in $r$.  
In particular, going to a basis labelled by the conformal weight, we can write the correspondence
\begin{align}
    h_{ij}(\lambda,x)
    = \frac{1}{\sqrt{2\pi}}\int_0^\infty \frac{dr}{r}\, r^{\Delta_- - 2}\,\mathscr{H}_{ij}(r,x),
\end{align}
where $\mathscr{H}_{ij}$ denotes the radial-gauge perturbation of the original metric $G_{IJ}$.  
Using the Weyl rescaling \eqref{Weylmetric} and it is related to $H_{ij}$ by
\begin{align}
    H_{ij}(\xi,x) = e^{-2\xi}\,\mathscr{H}_{ij}(\xi,x).
\end{align}
This is the counterpart of the formula \eqref{maxmiyamiya} in Maxwell theory.
Similarly, the counterpart of \eqref{maxmiya} can be written in the embedding
formalism as
\begin{align}
    h_{IJ}(\lambda,x)
    = \frac{1}{\sqrt{2\pi}}\int_0^\infty \frac{dr}{r}\, r^{\Delta_-}\,\mathscr{H}_{IJ}(r,x).
\end{align}
In analogy with \eqref{uplifting_maxwell}, we define
\begin{align}
    \frac{\partial x^I}{\partial x^i}\frac{\partial x^J}{\partial x^j}\,h_{IJ} := h_{ij},
\end{align}
and impose the transverse condition $x^I h_{IJ} = 0$.

\subsubsection{Pure gauge mode}

We consider a pure gauge perturbation $2 \partial_{(I}\zeta_{J)}$ in the flat $(D+2)$-dimensional space,
subject to the radial gauge condition $X^I \partial_{(I}\zeta_{J)}=0$.  
Equivalently, it satisfies
\begin{align}
  (X \cdot \partial)\,\zeta_I  - \zeta_I  + \partial_I(X \cdot\zeta)=0.
  \label{eq:radial-preserving-PDE}
\end{align}
Solving this, one finds the general residual diffeomorphism preserving the radial gauge,
\begin{align}
 \zeta&=f(x)\partial_r+\qty(\frac{1}{r}\nabla^i f(x)+W^i(x))\partial_i
  \label{eq:zeta-general}
\end{align}
where $f$ is a scalar on $S^{D+1}$ and $W^i$ is a vector field on $S^{D+1}$.

Following the method of \cite{Furugori:2025xkl}, we analytically continue $\mathbb{R}^{D+2}$ to $(D+2)$-dimensional Minkowski spacetime $\mathbb{M}^{D+2}$. 
We parameterize $\mathbb{R}^{D+2}$ by coordinates $(r,x)$ and introduce a polar angle $\theta$ via
\begin{align}
  x^{D+2} &= -\cos\theta, ~~~~~~~
  \bm{x} = (x^1,\ldots,x^{D+1}) = \sin\theta \,\hat n,
\end{align}
where $\hat n$ is a unit vector. 
We then perform the analytic continuation
\begin{align}
  \theta = \frac{\pi}{2} + i t \,,
  \label{eq:ac}
\end{align}
which maps $\mathbb{R}^{D+2}$ to a part of $\mathbb{M}^{D+2}$. 
For further details of this construction, see~\cite{Furugori:2025xkl}.

We will see that the diffeomorphism generated by the above $\zeta$ in \eqref{eq:zeta-general} contains the supertranslation and the superrotation after the continuation to Minkowski space. 
In particular, the scalar part associated with $f$ reproduces the supertranslation, and the vector one with $W^i$ does the superrotation by choosing specific $f$ and $W^i$. 
It enables us to find the counterparts in $(D+1)$-dim de Sitter space of these asymptotic symmetries through the ``dimensional reduction" procedure described above.
We can thus identify the dS counterparts of the Nambu-Goldstone modes for the asymptotic symmetries (supertranslation and superrotation).

\paragraph{Scalar part $f$: supertranslations}
Setting $W^i=0$ in \eqref{eq:zeta-general} gives
\begin{align}
    \zeta=f\partial_r+\frac{1}{r}\nabla^i f\partial_i.
    \label{eq:zeta-scalar}
\end{align}
We continue it to the Minkowski space using \eqref{eq:ac}.
The coordinates $(r,t)$ are Rindler–like ones in Minkowski space. To compare \eqref{eq:zeta-scalar} with the standard asymptotic symmetry near the future null infinity, we change the coordinates to the retarded Bondi coordinates $(U, R)$ as
\begin{align}
  U=-re^{-t},\qquad R=r\cosh t .
  \label{eq:UR-def}
\end{align}
Under the coordinate transformations, the vectors transform as
\begin{align}
  \partial_r \,=\, -e^{-t}\partial_U + \cosh t\,\partial_R,
  \qquad
  \partial_t \,=\, r e^{-t}\partial_U + r\sinh t\,\partial_R .
  \label{eq:drdt-to-UR}
\end{align}
Using \eqref{eq:drdt-to-UR} and treating $f=f(t,\Omega)$, one can show that, up to subleading $1/R$ terms, \eqref{eq:zeta-scalar} takes the form
\begin{align}
  \zeta  =  e^{-t} \left(-1-\partial_t\right)f \partial_U
             +  \big(f\cosh t - (\partial_t f)\sinh t\big)\partial_R
             +  \frac{1}{r\cosh^2 t}\,\nabla^\alpha f\,\partial_\alpha
             +  \mathcal{O} \left(R^{-2}\right).
  \label{eq:zeta-Bondi-pre}
\end{align}
To preserve the asymptotic structure of asymptotically flat spacetimes, the leading behavior of \(f\) must take the form
\begin{align}
  f(t, x^\alpha)
  = -\frac12\, e^{t}\, F(x^\alpha) + (\text{subleading}),
  \label{eq:f-choice}
\end{align}
Indeed, assuming this leading behavior, one recovers the standard supertranslation generator near future null infinity (large \(R\)).
\begin{align}
  \zeta
   =  F(x^\alpha)\,\partial_U
        + G(x^\alpha)\,\partial_R
        - \frac{U+2R}{2R^2}\,\nabla^\alpha F(x^\alpha)\,\partial_\alpha
        + \mathcal{O}(R^{-2}) .
  \label{eq:Bondi-supertranslation}
\end{align}
Here the coefficient $G(x^\alpha)$ of $\partial_R$ is partially determined by the leading contribution of $f$, which gives $-F(x^\alpha)$. However, it can be further modified by the subleading contribution of order $e^{-t}$, and we therefore denote the total coefficient, including this contribution, by $G(x^\alpha)$.
Indeed, the coefficient of $\partial_R$ in the supertranslation is known to be non-universal (i.e., gauge dependent) \cite{Bondi:1962px, Sachs:1962wk, Sachs:1962zza, Strominger:2017zoo}. Any choice of $G(x)$ does not affect the subsequent analysis.

This expression is consistent with \cite{Strominger:2017zoo} at leading order.
Re-expressing \eqref{eq:Bondi-supertranslation} back in the original $(r,t,\Omega)$ coordinates using \eqref{eq:UR-def} gives a compact radial–gauge form,
\begin{align}
  \zeta  &=  e^{t} \left(-\frac{F}{2}\,\partial_r  +  \frac{F}{2r}\,\partial_t  +  \frac{1}{r\cosh^2 t}\,\nabla^\alpha F\,\partial_\alpha\right)+ (\text{subleading})
  \nonumber\\
  &\sim  \frac{e^{t}}{2}F(x^\alpha)\left(-\partial_r+\frac{1}{r}\partial_t\right) \quad (t\to\infty).
  \label{eq:radial-supertranslation}
\end{align}

\paragraph{Vector part $W^i$: superrotations}
Similarly, setting $f=0$ in \eqref{eq:zeta-general} and continuing it to the Minkowski space, we obtain
\begin{align}
  \zeta  =  W^t\,\partial_t + \frac{1}{\cosh^2t}W^\alpha\,\partial_\alpha .
  \label{SR_zeta}
\end{align}
As in the scalar part, we can compare it to the asymptotic symmetry by using the Bondi coordinates.
By specifying a large $t$ behavior of $W^i$, we can recover the superrotation, as we obtained supertranslation in the scalar part above. 
Such $W^i$ are given by
\begin{align}
\begin{aligned}
  W^t  &=  - \frac1D\,\nabla_\alpha Y^\alpha  +  (\text{subleading}),\\
  \qquad
  W^\alpha  &= \frac{1}{4}e^{2t}Y^\alpha  +  (\text{subleading}),
  \label{eq:W-leading}
\end{aligned}
\end{align}
where $Y^\alpha$ is an arbitrary vector on $S^{D}$ as $Y \in \mathrm{Diff}(S^D)$.
This class of the asymptotic symmetry is called the generalized BMS symmetry \cite{Campiglia:2014yka}.\footnote{In $D=2$, we may restrict $Y$ to a local conformal Killing vector on $S^{2}$, satisfying 
\begin{align}
\nabla_{(\alpha}Y_{\beta)}  =  \frac{1}{D}\,\gamma_{\alpha\beta}\,\nabla \cdot Y.
\end{align}
Then, the asymptotic symmetry is the extended BMS symmetry \cite{Barnich:2011mi, Barnich:2009se}.}

Using \eqref{eq:drdt-to-UR}, one then finds that \eqref{SR_zeta} with \eqref{eq:W-leading} takes the following form,
\begin{align}
  \zeta  \sim  Y^\alpha\partial_\alpha
   - \frac{1}{D}(U+R)\,\nabla_\alpha Y^\alpha\,\partial_R
   + \frac{U}{D}\,\nabla_\alpha Y^\alpha\,\partial_U,
  \label{eq:Bondi-superrotation}
\end{align}
up to the subleading terms. 
The resulting expression is consistent with the \(D\)-dimensional \(\mathrm{Diff}(S^{D})\) asymptotic symmetry \cite{Capone:2021ouo, Capone:2023roc, Colferai:2020rte} when \(D\) is even.

\subsubsection{Gauge transformation in the sphere action}

The diffeomorphism generated by $\zeta$ acts on the metric perturbation $H_{ij}$ as
\begin{align}
 \delta\mathcal{H}_{ij}=2e^{\pm\qty(\frac{D}{2}-1)\xi}\qty(\nabla_i\nabla_j+\gamma_{ij})f+e^{\pm\frac{D}{2}\xi}\qty(\nabla_iW_j+\nabla_jW_i),
\end{align}
where $\nabla_i$ is the covariant derivative with respect to $\gamma_{ij}$.
The $\pm$ signs appearing in this equation arise, as in the Maxwell case, from the symmetry under $\xi \to -\xi$.
Performing the Fourier transform along $\xi$ with spectral parameter $\lambda$, we obtain  
\begin{align}
    \delta h_{ij}&=2\qty(\nabla_i\nabla_j+\gamma_{ij})f~\delta^G\qty(\lambda\pm i\qty(\frac{D}{2}-1))+\qty(\nabla_iW_j+ \nabla_jW_i)\delta^G\qty(\lambda\pm i\frac{D}{2}).
\end{align}
The first term corresponds, on $\mathrm{dS}_{D+1}$ slices, to the gauge transformation of the partially massless modes
$\delta h_{ij}\sim (\nabla_i \nabla_j+\gamma_{ij})\,f$.\footnote{
We review the gauge transformation of the partially massless graviton in App. \ref{FPreview}.
} 
The second term corresponds to the gauge transformation of the massless modes generated by a transverse vector on $\mathrm{dS}_{D+1}$ after the analytic continuation.

\section{Cosmological-Celestial Operator Correspondence}
\label{Sec.4}
In this section, we extend the correspondence between cosmological operators, which are extrapolated operators in de Sitter space, and celestial CFT operators, established for scalar fields in \cite{Furugori:2025xkl}, to fields with spin.
For simplicity, we first focus on a spin-1 field.

As a preparation, we would like to write cosmological spin-1 operators using the embedding formalism, so that the discussion can proceed in the same manner as in the scalar case.
First, the cosmological operators in de~Sitter space are defined as the leading coefficients of the bulk field in the late/early time limits $t\to \pm \infty$ (where our parametrization is given in \eqref{param_dS}). The angular components $a_{\alpha,\lambda}(t,\bm{p}(x^\alpha))$ of the gauge field on de~Sitter have the following asymptotic behavior (see, e.g., \cite{Baumann:2020dch})
\begin{align}
a_{\alpha,\lambda}(t,\bm{p}(x^\alpha))
\;\sim\; e^{\mp(\Delta_- -1)t}\,V_{\alpha,\Delta_-}(x^\alpha)
      + e^{\mp(\Delta_+ -1)t}\,W_{\alpha,\Delta_+}(x^\alpha)
\qquad (t\rightarrow \pm\infty),
\end{align}
where $x^\alpha$ are coordinates on the celestial sphere.

We uplift this scaling behavior to the embedding space  
using our parametrization of the de~Sitter embedding coordinates
\begin{align}
x^I(t,\bm{p}) = (\sinh t,\;\cosh t\,\bm{p}),
\qquad
p^I = (1,\bm{p}).
\end{align}
We also we define the operators \(V_I,W_I\) in the embedding space, following \cite{Simmons-Duffin:2012juh}, as\footnote{Note that the definition of the $V_{I,\Delta}(p),W_{I,\Delta}(p)$ in the embedding space has ambiguity: one may shift
\begin{align}
V_{I,\Delta}(p) &\;\to\; V_{I,\Delta}(p) + \beta(p)\,p_I,\\
W_{I,\Delta}(p) &\;\to\; W_{I,\Delta}(p) + \beta(p)\,p_I,
\end{align}
for an arbitrary scalar function $\beta(p)$. This reflects the standard null-cone redundancy of the embedding formalism for spinning primaries \cite{Simmons-Duffin:2012juh}.}
\begin{align}
\frac{\partial p^I}{\partial x^\alpha}V_{I,\Delta}(p):= V_{\alpha,\Delta}(p)\qquad \frac{\partial p^I}{\partial x^\alpha}W_{I,\Delta}(p):= W_{\alpha,\Delta}(p).
\end{align}
Using \eqref{uplifting_maxwell}, we then obtain
\begin{align}
\frac{\partial x^I}{\partial x^\alpha}\,a_{I,\lambda}(x)
\;\sim\;
e^{\mp(\Delta_- -1)t}\,\frac{\partial p^I}{\partial x^\alpha}V_{I,\Delta_-}(p)
+ e^{\mp(\Delta_+ -1)t}\,\frac{\partial p^I}{\partial x^\alpha}W_{I,\Delta_+}(p)
\qquad (t\rightarrow \pm\infty).
\end{align}
We can equivalently write it as 
\begin{align}
e^{\pm t}\,\frac{\partial p^I}{\partial x^\alpha}a_{I,\lambda}(x)
\;\sim\;
e^{\mp(\Delta_- -1)t}\,\frac{\partial p^I}{\partial x^\alpha}V_{I,\Delta_-}(p)
+ e^{\mp(\Delta_+ -1)t}\,\frac{\partial p^I}{\partial x^\alpha}W_{I,\Delta_+}(p)
\qquad (t\rightarrow \pm\infty),
\end{align}
where we have used
\begin{align}
\frac{\partial x^I}{\partial x^\alpha}\to e^{\pm t}\frac{\partial p^I}{\partial x^\alpha}\qquad (t\rightarrow \pm\infty).
\end{align}
Finally, stripping off the Jacobian factor and absorbing any overall normalization into the definition of the operators, we arrive at the embedding-space asymptotics
\begin{align}
\label{asympt_behavior_dS_field}
a_{I,\lambda}(t,\bm{p})
\;\sim\;
e^{\mp\Delta_- t}\,V_{I,\Delta_-}(p)
+ e^{\mp\Delta_+ t}\,W_{I,\Delta_+}(p)
\qquad (t\rightarrow \pm\infty).
\end{align}

To make contact with the celestial dictionary, we now examine the late-time limit $t\rightarrow +\infty$.
In this asymptotic region of the Minwkoski spacetime, we have the plane-wave expansion of the gauge field as
\begin{align}
A_I(X)
= \sum_{\alpha}\int \frac{\omega^{D-1}\,d\omega\,d^Dk}{2(2\pi)^{D+1}}
\left(a^{\alpha}(\omega k)\,f^{\alpha}_I(X)
     + a^{\alpha\dagger}(\omega {k})\,f^{\alpha\ast}_I(X)\right),
\end{align}
where $\alpha$ are the labels for the polarization.
Inserting it into \eqref{maxmiya}, we obtain
\begin{align}
a_{I,\lambda}(x)
= \frac{1}{\sqrt{2\pi}}\sum_{\alpha}\int \frac{dr}{r}\,r^{\Delta}
  \int \frac{\omega^{D-1}\,d\omega\,d^D{k}}{2(2\pi)^{D+1}}
  \left(a^{\alpha}(\omega {k})\,f^{\alpha}_I(X)
       + a^{\alpha\dagger}(\omega {k})\,f^{\alpha\ast}_I(X)\right).
\label{eq:spin1-free-expansion}
\end{align}
The mode functions $f^{\alpha}_I(X)$ are given by
\begin{align}
f^{\alpha}_I(X)
&=\left(\varepsilon^{\alpha}_I(k)
   -\frac{\varepsilon^{\alpha}(k)\cdot x}{k\cdot x}\,k_I\right)e^{i\omega k\cdot X}
 \;=\;\varepsilon^{\prime \alpha}_I(k)\,e^{i\omega k\cdot X},
\end{align}
where $\varepsilon^{\alpha}_I(k) := \partial^\alpha p_I(x^\alpha)$ are polarization vectors,
and we have the radial gauge condition $\varepsilon'(k)\cdot x = 0$.
 In the asymptotic region $t\rightarrow \pm\infty$, it is convenient to rewrite the polarization as
\begin{align}
f^{\alpha}_I(X)
&=\left(\varepsilon^{\alpha}_I(k)
   -\frac{\varepsilon^{\alpha}(k)\cdot p_{\pm}}{k\cdot p_{\pm}}\,k_I\right)e^{i\omega k\cdot X}
 \;=\;\varepsilon^{\prime \alpha}_I(k,p_{\pm})\,e^{i\omega k\cdot X},
\end{align}
where $p_{\pm} = (1,\pm\bm{p})$ are null vectors characterizing the asymptotic direction.\footnote{As $t \rightarrow \infty$ and $k$ approaches $p_{\pm}$, the polarization $\varepsilon^\prime$ chosen above becomes singular.
In this regime, one must switch to a different gauge choice.
For example, one may take the polarization vector $\varepsilon'$ by
\begin{align}
  \varepsilon'_I(k):=\varepsilon^{\alpha}_{I}(k) = \partial^{\alpha} p_{I}(x^{\alpha}) \, .
\end{align}
}

We now define the polarization-contracted annihilation operator
\begin{align}
a_I(\omega \bm{k})
:= \sum_\alpha \varepsilon^{\prime\alpha}_I(k,p_{\pm})\,a^{\alpha}(\omega {k}),
\end{align}
and similarly for the creation operator. Substituting this into \eqref{eq:spin1-free-expansion}, we obtain
\begin{align}
a_{I,\lambda}(x)
= \frac{1}{\sqrt{2\pi}}\int \frac{dr}{r}\,r^{\Delta}
  \int \frac{\omega^{D-1}\,d\omega\,d^Dk}{2(2\pi)^{D+1}}
  \left(a_{I}(\omega k)\,e^{i\omega k\cdot X}
       + a_{I}^{\dagger}(\omega k)\,e^{-i\omega k\cdot X}\right).
\end{align}

As done in \cite{Furugori:2025xkl}, considering the asymptotic limit, using the celestial dictionary \eqref{celestial_scr}, 
performing the Mellin transform over $\omega$, we can identify the leading coefficients $V_{I,\Delta_-}(p)$ and $W_{I,\Delta_+}(p)$ in the de~Sitter asymptotics \eqref{asympt_behavior_dS_field}  with the spin-1 celestial operators and their shadows.
More precisely, up to overall normalization and possible phases, the cosmological operators are related to the celestial ones as\footnote{
More precisely, it should be written as
\begin{align}
V_{I,\Delta_-}(p)
&\propto
   i^{-\Delta_-}\,\widetilde{\mathscr{O}_{I,\Delta_+}}(p) + i^{\Delta_-}\,\bigl(\widetilde{\mathscr{O}_{I,\Delta_-}}(p)\bigr)^\dagger,
\end{align}
and
\begin{align}
W_{I,\Delta_+}(p)
&\propto
   i^{-\Delta_+}\,\mathscr{O}_{I,\Delta_+}(p) + i^{\Delta_+}\,\bigl(\mathscr{O}_{I,\Delta_-}(p)\bigr)^\dagger.
\end{align}
However, in this paper, we omit the latter term, since it contributes only when we consider two-point functions.
}
\begin{align} \label{shadow_primary}
V_{I,\Delta_-}(p)
&\propto
   i^{-\Delta_-}\,\widetilde{\mathscr{O}_{I,\Delta_+}}(p),
\end{align}
and
\begin{align} \label{shadow_primary2}
W_{I,\Delta_+}(p)
&\propto
   i^{-\Delta_+}\,\mathscr{O}_{I,\Delta_+}(p).
\end{align}
Here $\mathscr{O}_{I,\Delta_\pm}(p)$ denote spin-1 celestial primaries of dimensions $\Delta_\pm$, while $\widetilde{\mathscr{O}}$ denotes their shadow transforms. \eqref{shadow_primary} and \eqref{shadow_primary2} thus provide the operator dictionary relating de~Sitter cosmological operators for spin-1 fields to spin-1 operators in the celestial CFT.

We can obtain similar expressions in the case of spin-2 as follows:
\begin{align}
V_{IJ,\Delta_-}(p)
&\propto
   i^{-\Delta_-}\,\widetilde{\mathscr{O}_{IJ,\Delta_+}}(p),
\end{align}
and
\begin{align}
W_{IJ,\Delta_+}(p)
&\propto
   i^{-\Delta_+}\,\mathscr{O}_{IJ,\Delta_+}(p).
\end{align}

\section{Conformally Soft Theorem from Ward-Takahashi Identity of Cosmological Correlators}\label{Sec.5}
In this section, we use the operator dictionary obtained in Sec.~\ref{Sec.4} to examine the equivalence between the Ward-Takahashi identities of cosmological correlators and the conformally soft theorem in celestial holography. In Sec.~\ref{Sec.3}, we employed a dimensional reduction procedure to relate fields in Minkowski space to fields on de Sitter space. 
There, we showed that gauge transformations in Minkowski space can be interpreted as gauge symmetries acting on the massless and partially massless modes on de Sitter space. 
From the perspective of the cosmological correlators \cite{Baumann:2020dch}, these massless modes are dual to the conserved currents associated with these asymptotic symmetries. 
We thus have the Ward-Takahashi identities of cosmological correlators associated with the current conservation.
We will show that these Ward-Takahashi identities are equivalent to the celestial soft theorem for the corresponding gauge-field modes.

\subsection{Maxwell case: Leading}
As a simple toy model, let us consider a $U(1)$ gauge theory in which a gauge field $a_I$ is coupled to a single massless charged scalar field with charge $q$  in $\mathbb{R}^{D+2}$.
After the mapping to dS, we should have the Ward-Takahashi identity associated with the $U(1)$ symmetry for extrapolated cosmological operators as \cite{Baumann:2020dch}
\begin{align}
    \ev{D_{I}J^{I}(p)\prod_{i}{O}_{\Delta_i}(p_i)}=\sum_i\delta(p-p_i)\ev{\delta{O}_{\Delta_i}(p_i)\prod_{j\neq i}{O}_{\Delta_j}(p_j)}
    \label{u1dsward}
\end{align}
where $J^I$ is the conserved current associated with the $U(1)$ symmetry and $O_{\Delta}$ are extrapolated operators for scalar modes $\psi_{\lambda}$ in dS, where $\psi_{\lambda}$ is the dS scalar fields obtained from the scalar field in  Minkowski space.

From \eqref{eq:pure-gauge-delta}, we saw that the conformal dimensions corresponding to the $U(1)$ pure-gauge modes are given by $\lambda = \pm i\frac{D-2}{2}$.
It then follows that the associated cosmological gauge-field mode can be expanded as
\begin{align}
\begin{aligned}
    a_{I,\lambda=i\frac{D-2}{2}}\sim e^{-(D-1) t}V_{I, \Delta=D-1}(p)+e^{-t}{W}_{I,\Delta=1}(p),\\
    a_{I,\lambda=-i\frac{D-2}{2}}\sim e^{- t}V_{I,\Delta=1}(p)+e^{-(D-1)t}{W}_{I,\Delta=D-1}(p).
\end{aligned}
\end{align}
In $D=2$, these two asymptotic behaviors become degenerate, and one finds $V_{\Delta=1} = W_{\Delta=1}$. In order to obtain a complete basis of modes, it is therefore necessary to take an appropriate linear recombination. This phenomenon is generic in (anti-)de Sitter spacetime \cite{Henneaux:2004zi, Perez:2015jxn}, and an analogous procedure is also required in celestial holography \cite{Donnay:2018neh}. However, our present goal is not to construct a complete basis of modes but simply to identify the relevant current. Since the additional mode does not play any role in this identification, the degeneracy is not an essential issue.

Similarly, the asymptotic behavior of the scalar field is written as
\begin{align}
    \psi_{\lambda}\sim e^{-\Delta_- t}V_{\Delta_-} + e^{-\Delta_+ t}W_{\Delta_+}.
\end{align}
The action of the gauge transformation on these cosmological operators can then be expressed as follows
\begin{align}
    \delta O_{\Delta}\sim iqO_{\Delta},
\end{align}
where $O_\Delta = V_{\Delta}, W_\Delta$.
Moreover, the spin-1 cosmological operators are related to the celestial operators via
\begin{align}
\begin{aligned}
    W_{I,\Delta_+}(p)
&\propto
   i^{-\Delta_+}\mathscr{O}_{I,\Delta_+}(p),\\
   V_{I,\Delta_-}(p)
&\propto i^{-\Delta_-}\widetilde{\mathscr{O}_{I,\Delta_+}}(p).
\end{aligned}
\end{align}
In particular, we identify\footnote{In this setting, $W_{\Delta=D-1}$ may also be served as a $U(1)$ current, based solely on its conformal dimension. The issue probably arises because we have not specified the degrees of freedom in the asymptotic boundary (edge modes) in Minkowski space. We leave this issue for future work, and in this paper, we simply adopt the prescription that the operator $V_{D-1}$ plays the role of the current.} the $U(1)$ current with the $\Delta = D-1$ mode as\footnote{
Strictly speaking, in celestial holography for $D=2$, the definition of the current requires multiplying the operator by a prefactor $\Delta-1$ and then
taking the limit $\Delta \to 1$.  
This procedure cancels the divergence in the normalization factor of the
celestial dictionary and yields a finite result.  
In the present discussion, we will not keep track of this overall
prefactor.
}
\begin{align}
    J^{I}= V^{I}_{\Delta=D-1}\sim i^{-(D-1)}\widetilde{\mathscr{O}_{\Delta=1}^I}.
\end{align}

Using this correspondence, the Ward-Takahashi identity \eqref{u1dsward} can be rewritten in terms of celestial operators as
\begin{align}
\ev{D_{I}\widetilde{\mathscr{O}^I_{\Delta=1}}(p)\prod_{i}\mathscr{O}_{\Delta_i}}\propto\sum_i q\delta(p-p_i)\ev{\prod_{ i}\mathscr{O}_{\Delta_i}}.
\end{align}
This equation is precisely the conformally soft photon theorem \cite{Pano:2023slc, Kapec:2021eug}. 
It therefore establishes the equivalence between the Ward-Takahashi identity of the cosmological correlators with $U(1)$ current and the conformally soft photon theorem.

Especially for $D=2$, the shadow and non-shadow operators at scaling dimension $\Delta = 1$ become degenerate.
Working on the Poincar\'e section of the celestial sphere, which we identify with the complex plane $\mathbb{C}$ endowed with coordinates $(z,\bar z)$ and flat metric $\gamma_{z\bar z}=\gamma_{\bar z z}=1$\footnote{
The Poincaré section is obtained by parametrizing a null vector $p^I$ in terms of $(z,\bar z)$ as
\begin{align}
p^I(z,\bar z)
=
\bigl(1+z\bar z,\;z+\bar z,\;-i(z-\bar z),\;1-z\bar z\bigr).
\end{align}
}, we obtain
\begin{align}
    \ev{\mathscr{O}_{z,\Delta=1}(z,\bar z)\prod_{i}\mathscr{O}_{\Delta_i}(z_i,\bar z_i)}\propto\sum_i \frac{q}{z-z_i}\ev{\prod_{ i}\mathscr{O}_{\Delta_i}(z_i, \bar z_i)},
\end{align}
where we have used the following distributional identity:
\begin{align}
\partial_{\bar z} \left(\frac{1}{z-w}\right) = 2\pi\,\delta^{(2)}(z-w).
\label{disfomula}
\end{align}
This is consistent with the celestial Ward identity in $D=2$ \cite{He:2019jjk}.

\subsection{Gravity case: Leading}

We now derive the leading conformally soft graviton theorem from the partially massless gauge symmetry in de~Sitter space. We begin with the Ward-Takahashi identity of the cosomological correlators including  the extrapolated partially massless current $J^{IJ}$ associated with the partially massless modes,
\begin{align}
    \Big\langle D_{I}D_{J} J^{IJ}(p)\,\prod_{i} O_{\Delta_i}(p_i)\Big\rangle
    \;=\;
    \sum_i \delta(p-p_i)\,
    \Big\langle \delta O_{\Delta_i}(p_i)\,\prod_{j\neq i} O_{\Delta_j}(p_j)\Big\rangle,
    \label{eq:dS-PM-Ward}
\end{align}
where $\delta O_{\Delta_i}$ denotes the variation of the scalar extrapolated operator induced by the partially massless gauge transformation. The partially massless graviton modes in de~Sitter admit the late-time expansion
\begin{align}
    h_{IJ,\lambda =i\frac{D-2}{2}}
    &\sim e^{-(D-1)t}\,V_{IJ, \Delta = D-1}(p)
        + e^{-t}\,W_{IJ, \Delta = 1}(p).
    \label{eq:h-PM-expansion}
\end{align}

To extract the action of the partially massless symmetry on extrapolated operators, we first consider its effect on a bulk scalar field $\Phi$. 
In sec.~\ref{subsec:2gravity}, we saw that the gauge transformation for the partially massless modes in dS is associated with a diffeomorphism in the Minkowski space given by \eqref{eq:radial-supertranslation}.
Let $F(p)$ be the scalar gauge parameter on the celestial sphere $S^D$. In the asymptotic region near future infinity $t\sim \infty$, the diffeomorphism induces 
\begin{align}
    \delta \Phi
    \;\sim\;
    \frac{1}{2}\,F\,e^{t}\Bigl(-\partial_r+\frac{1}{r}\partial_t\Bigr)\Phi
    + {O}(e^{-t}).
    \label{eq:delta-phi-PM}
\end{align}
We then consider the scalar modes $\psi_\lambda(x)$ in dS associated with $\Phi$ \cite{Furugori:2025xkl}
\begin{align}
    \psi_\lambda(x)
    \;=\;
    \frac{1}{\sqrt{2\pi}}\int_0^\infty dr\,r^{\Delta_- - 1}\,\Phi(r,x). 
\end{align}
Using \eqref{eq:delta-phi-PM}, we obtain at leading order in the late-time expansion
\begin{align}
    \delta\psi_\lambda(x)
    &\sim \frac{1}{\sqrt{2\pi}}\int_0^\infty dr\,r^{\Delta_--1}\delta\Phi\nonumber\\
    &\sim \frac{1}{2\sqrt{2\pi}}\,F e^t \int_0^\infty dr\,r^{\Delta_--1}\Bigl(-\partial_r+\frac{1}{r}\partial_t\Bigr)\Phi
    \nonumber\\
    &\sim \frac{1}{2}\,F e^t\,\bigl(\Delta_- - 1 + \partial_t\bigr)\psi_{\lambda - i} + \mathcal{O}(e^{-t}),
    \label{eq:delta-psi-PM}
\end{align}
where $\psi_{\lambda-i}$ arises because the partially massless shift effectively changes the radial exponent by one unit. The late-time behavior of $\psi_{\lambda-i}$ can be parametrized as
\begin{align}
    \psi_{\lambda-i}(x)
    \;\sim\;
    e^{-(\Delta_- - 1)t}\,V_{\Delta_- - 1}(p)
    + e^{-(\Delta_+ + 1)t}\,W_{\Delta_+ + 1}(p)
    + \cdots,
    \label{eq:psi-lambda-i-expansion}
\end{align}
where the ellipsis denotes subleading terms in $e^{-2t}$. Substituting \eqref{eq:psi-lambda-i-expansion} into \eqref{eq:delta-psi-PM} and matching the powers of $e^{-t}$, we see that $\delta W_{\Delta_+}$ is given by 
\begin{align}
    \delta W_{\Delta_+}
    \;&\sim\;
    \frac{1}{2}\bigl(\Delta_- - \Delta_+ - 2\bigr)\,W_{\Delta_+ + 1}\nonumber\\
    \;&\sim\;
    -(1 + i\lambda)\,W_{\Delta_+ + 1}.
    \label{eq:delta-O-plus}
\end{align}
Thus, the partially massless symmetry acts non-trivially on the ``plus'' branch like this, shifting its conformal dimension by one unit (weight shift).

On the other hand, for $\delta V_{\Delta_-}$, 
the leading contribution proportional to $e^{-(\Delta_- - 1)t}$ vanishes because
\begin{align}
    \bigl(\Delta_- - 1 + \partial_t\bigr)e^{-(\Delta_- - 1)t} = 0.
    \label{subV}
\end{align}
The leading term of $\delta V_{\Delta_-}$ is determined by the subleading $O(e^{-t})$ contribution to $\delta\Phi$ together with the subleading terms in the asymptotic expansion of $\psi_{\lambda-i}(x)$,
\begin{align}
    \psi_{\lambda-i}(x)
    \sim e^{-(\Delta_- -1)t}\Bigl(V_{\Delta_- -1}
        + e^{-2t} V^{(2)}_{\Delta_- -1} + \cdots\Bigr)
      + e^{-(\Delta_+ +1)t}\Bigl(W_{\Delta_+ +1}
         + \cdots\Bigr).
\end{align}
The subleading term in \eqref{eq:delta-phi-PM} includes a term $\frac{e^{-t}}{r }\,\nabla^\alpha F\,\nabla_\alpha \Phi$. One sees that, after integration by parts, it acquires two derivatives. Moreover, by analogy with the Fefferman--Graham expansion, $V^{(2)}_{\Delta_- -1}$ can be interpreted as the level-two descendant operator of $V_{\Delta_- -1}$ \cite{deHaro:2000vlm, Skenderis:2002wp, Freedman:1998tz}. It therefore follows that the conformal dimension of $\delta V_{\Delta_-}$ is $\Delta_- + 1$.
This coincides with the conformal dimension of the operator obtained by first acting with a supertranslation (weight shift) on $\mathscr{O}_{\Delta_+}$ and subsequently applying the shadow transformation.

We now use the cosmological-celestial dictionary to translate \eqref{eq:dS-PM-Ward} into the conformally soft graviton theorem. In analogy with the $U(1)$ case, the partially massless current at the boundary is associated with the $\Delta=D-1$ graviton mode and is mapped to the shadow of the spin-$2$ conformal primary in celestial CFT. We can identify the partially massless current with the $\Delta = D-1$ mode as
\begin{align}
    J^{IJ}= V^{IJ}_{\Delta=D-1}\sim -i\widetilde{\mathscr{O}_{\Delta=1}^{IJ}}.
\end{align}
Denoting the corresponding celestial operator by $\widetilde{\mathscr{O}^{IJ}_{\Delta=1}}(p)$ and using the normalization of \eqref{eq:delta-O-plus}, the Ward-Takahashi identity \eqref{eq:dS-PM-Ward} becomes
\begin{align}
    \Big\langle D_{I}D_{J}\,\widetilde{\mathscr{O}^{IJ}_{\Delta=1}}(p)\,\prod_{i}\mathscr{O}_{\Delta_i}(p_i)\Big\rangle
    \propto
    \sum_i \delta(p-p_i)\,
    \Big\langle \mathscr{O}_{\Delta_i+1}(p_i)\prod_{j\neq i}\mathscr{O}_{\Delta_j}(p_j)\Big\rangle.
    \label{eq:celestial-PM-shadow}
\end{align}
This equation is precisely the celestial Ward-Takahashi identity \cite{Pano:2023slc, He:2019jjk}.
 Finally, using the Mellin representation that relates momentum-space amplitudes to celestial correlators, the insertion of $\mathscr{O}_{\Delta_i+1}(p_i)$ can be written as the action of an energy operator $\hat{\omega}_i$ on the celestial correlator,\footnote{
In our conventions, celestial operators are defined by a Mellin transform
\begin{align}
  \mathscr{O}_\Delta(p)
  \;=\;
  \int_0^\infty d\omega\,\omega^{\Delta-1}\,a(p)\,,
\end{align}
so that multiplying the momentum-space amplitude by the energy
$\omega$,
is equivalent to a unit shift of the conformal weight,
$\Delta\to\Delta+1$. 
We write this weight-shifting operator as $\hat{\omega}$.
}
\begin{align}
    \Big\langle \mathscr{O}_{\Delta_i+1}(p_i)\prod_{j\neq i}\mathscr{O}_{\Delta_j}(p_j)\Big\rangle
    \;=\;
    \hat{\omega}_i\,\Big\langle\prod_{j}\mathscr{O}_{\Delta_j}(p_j)\Big\rangle.
\end{align}
Especially for $D=2$, using the fact that shadow and non-shadow operators are degenerate \cite{Donnay:2018neh}, we obtain, after partial integration of \eqref{eq:celestial-PM-shadow},
\begin{align}
    \Big\langle \,\mathscr{O}_{zz,\,\Delta=1}(z,\bar z)\,\prod_{i}\mathscr{O}_{\Delta_i}(z_i,\bar z_i)\Big\rangle
    \propto
    \sum_i \frac{\bar z-\bar z_i}{z-z_i}\hat{\omega}_i
    \Big\langle\prod_{j}\mathscr{O}_{\Delta_j}(z_i,\bar z_i)\Big\rangle.
\end{align}
This is consistent with the celestial Ward-Takahashi identity for the supertranslation (which is equivalent to the leading soft graviton theorem) for $D=2$ \cite{He:2014laa, Strominger:2013jfa}.

\subsection{Gravity case: Subleading}
We now turn to derive the celestial Ward-Takahashi identity associated with the subleading soft graviton from diffeomorphism symmetry in de~Sitter space. 
In this section, we focus only on the $D=2$ case.

We start from the Ward-Takahashi identity of the cosmological correlators including the stress-tensor dual to the massless graviton in de~Sitter space,
\begin{align}
    \int [d^2 p]\; Y_J(p)\,
    \Big\langle D_{I} T^{IJ}(p)\prod_{i} O_{\Delta_i}(p_i)\Big\rangle
    \;=\;
    \int [d^2 p]\;\sum_i \delta(p-p_i)\,
    \Big\langle \delta_Y O_{\Delta_i}(p_i)\prod_{j\neq i} O_{\Delta_j}(p_j)\Big\rangle,
    \label{eq:dS-grav-Ward}
\end{align}
where $T^{IJ}$ is the stress-tesnor dual to the massless graviton, $Y^I$ is a vector on $S^2$ (superrotation parameters) in the embedding formalism, and $O_{\Delta_i}$ are scalar operators. The massless gravitons in de~Sitter admit the late-time expansion
\begin{align}
   h_{IJ,\lambda = i}
    &\sim e^{-2 t}V_{IJ,\Delta = 2}(p) +  W_{IJ, \Delta = 0}(p),
    \label{eq:hIJ-expansion}
\end{align}
and $V_{IJ,\Delta = 2}$ is the extrapolated operator for the massless graviton modes, while $W_{IJ, \Delta = 0}$ plays the role of the shadow partner.

Next, we determine how the diffeomorphism generated by $Y^I$ acts on scalar fields. Writing the bulk diffeomorphism vector as $W^a$ and decomposing it into the time and angular components on de~Sitter (see \eqref{SR_zeta}), 
\begin{align}
    \delta_W \phi
    \;=\; W^a \partial_a \phi
    \;=\; \qty(-W^t \partial_t + \frac{1}{\cosh^2 t} W^\alpha \partial_\alpha)\phi,
\end{align}
one finds, using the late-time asymptotics $W^t \sim -\frac{1}{2} \nabla_\alpha Y^\alpha$ and $W^\alpha \sim \frac{1}{4}e^{2t}Y^\alpha$, that
\begin{align}
    \delta_W \phi
    \;\sim\;
    \Bigl(\frac{1}{2} \nabla_\alpha Y^\alpha \,\partial_t +  Y^\alpha \partial_\alpha\Bigr)\phi
    \qquad (t\to +\infty).
\end{align}
Then, the cosmological modes $\psi_\lambda$ associated with $\phi$ transform as 
\begin{align}
    \delta \psi_\lambda(x)
    \;\sim\;
    \Bigl(\frac{1}{2} \nabla_\alpha Y^\alpha \,\partial_t + Y^\alpha \partial_\alpha\Bigr)\psi_\lambda(x).
\end{align}
Expanding the cosmological mode near future infinity as
\begin{align}
    \psi_\lambda
    \;\sim\;
    e^{-\Delta_- t} V_{\Delta_-}(p)
    + e^{-\Delta_+ t} W_{\Delta_+}(p),
\end{align}
and matching powers of $e^{-\Delta t}$, we can find that the corresponding extrapolated operators $O_\Delta$ transform as
\begin{align}
    \delta_Y O_{\Delta}(p)
    \;\sim\;
    \Bigl(-\frac{\Delta}{2}\,\nabla_\alpha Y^\alpha(p)
          + Y^\alpha(p)\,\partial_\alpha\Bigr) O_{\Delta}(p).
    \label{eq:scalar-diffeo-transform}
\end{align}

Recall that the massless graviton in de~Sitter is associated with the $V_{IJ,\Delta = 2}$ in \eqref{eq:hIJ-expansion}. Then, in analogy with the $U(1)$ case, we identify
\begin{align}
    T^{IJ}
    \;\sim\;
    V^{IJ}_{\Delta = 2}(p)
    \;\propto\;-\widetilde{{\mathscr{O}^{IJ}_{\Delta = 0}}}(p)\equiv-{\mathscr{T}^{IJ}_{\Delta = 2}}(p),
\end{align}
where $\mathscr{T}^{IJ}_{\Delta = 2}(p)$ is the spin-2 celestial operator (the celestial stress tensor) \cite{Kapec:2016jld, Donnay:2020guq}. 

We now work on the Poincar\'e section of the celestial sphere. Substituting the transformation law \eqref{eq:scalar-diffeo-transform} into the right-hand side of \eqref{eq:dS-grav-Ward}, and rewriting the left-hand side in terms of ${\mathscr{T}}^{IJ}_{\Delta = 2}$ using the above dictionary, we obtain the celestial Ward-Takahashi identity
\begin{align}
\int_{\mathbb{C}} d^2 z\; Y_\beta(z,\bar z)\,
\Big\langle \partial_\alpha\,\mathscr{T}^{\,\alpha\beta}_{\Delta=2}(z,\bar z)\;\prod_{i=1}^n\mathscr{O}_{\Delta_i}(z_i,\bar z_i)\Big\rangle
= \sum_{i=1}^n\qty[\frac{\Delta_i}{2}\,\partial_\alpha Y^\alpha(z_i,\bar z_i)-Y^\alpha(z_i,\bar z_i)\,\partial_{z_i^\alpha}]\,
\Big\langle \prod_{j=1}^n\mathscr{O}_{\Delta_j} \Big\rangle.
\label{eq:celestial-grav-Ward}
\end{align}
Using integration by parts, the left-hand side of \eqref{eq:celestial-grav-Ward} can be rewritten as
\begin{align}
\int_{\mathbb{C}} d^2 z\; Y_\beta\,
\Big\langle \partial_\alpha\,\mathscr{T}^{\alpha\beta}_{\Delta=2}\prod_{i=1}^n\mathscr{O}_{\Delta_i}\Big\rangle
= - \int_{\mathbb{C}} d^2 z\; (\partial_\alpha Y_\beta)\,
\Big\langle \mathscr{T}^{\alpha\beta}_{\Delta=2}\prod_{i=1}^n\mathscr{O}_{\Delta_i}\Big\rangle .
\end{align}
On the plane, the only non-vanishing components of the spin-2 celestial operator are the holomorphic and anti-holomorphic ones,
\begin{align}
T_{\text{cel}}(z) := 2\pi\mathscr{T}_{zz,\,\Delta=2}(z,\bar z), 
\qquad
\bar T_{\text{cel}}(\bar z) := 2\pi\mathscr{T}_{\bar z\bar z,\,\Delta=2}(z,\bar z),
\end{align}
with $\mathscr{T}^{\,z\bar z}_{\Delta=2}=\mathscr{T}^{\,\bar z z}_{\Delta=2}=0$. Using $Y_z = Y^{\bar z}$ and $Y_{\bar z}=Y^z$, the contraction becomes
\begin{align}
2\pi(\partial_\alpha Y_\beta)\,\mathscr{T}^{\alpha\beta}_{\Delta=2}
= (\partial_z Y^{\bar z})\,\bar T_{\text{cel}}(\bar z)
+ (\partial_{\bar z}Y^z)\, T_{\text{cel}}( z).
\end{align}
To extract the local celestial Ward-Takahashi identity for the holomorphic stress tensor, we choose a CKV that probes only the $z$-sector,
\begin{align}
Y^{\bar z}(z,\bar z) = 0, 
\qquad
Y^z(z,\bar z) = \frac{1}{z-w},
\end{align}
where $\varepsilon(z)$ is a holomorphic test function and $w\in\mathbb{C}$ is the insertion point. With this choice $\partial_\alpha Y^\alpha = \partial_z Y^z$, the identity reduces to
\begin{align}
- \int_{\mathbb{C}} d^2 z\; \frac{1}{2\pi}(\partial_{\bar z}Y^z)\,
\big\langle  T_{\text{cel}}(z)\prod_{i=1}^n\mathscr{O}_{\Delta_i}\big\rangle
= \sum_{i=1}^n\qty[\frac{\Delta_i}{2}\,\partial_z Y^z(z_i)
- Y^z(z_i)\,\partial_{z_i}]\Big\langle \prod_{j=1}^n\mathscr{O}_{\Delta_j} \Big\rangle.
\end{align}
Using \eqref{disfomula} and writing the scalar celestial primaries as $\mathscr{O}_{\Delta_i}(z_i,\bar z_i)$ with conformal weights $(h_i,\bar h_i)=(\Delta_i/2,\Delta_i/2)$, we arrive at
\begin{align}
\big\langle  T_{\text{cel}}( w)\prod_{i=1}^n\mathscr{O}_{\Delta_i}(z_i,\bar z_i)\big\rangle
= \sum_{i=1}^n\left[\frac{\Delta_i}{2}\frac{1}{(w-z_i)^2}
+\frac{1}{w-z_i}\,\partial_{z_i}\right]
\Big\langle \prod_{j=1}^n\mathscr{O}_{\Delta_j}(z_j,\bar z_j) \Big\rangle,
\label{eq:celestial-Ward-final}
\end{align}
together with the anti-holomorphic counterpart obtained by $z\leftrightarrow\bar z$, $T_{\text{cel}}\leftrightarrow\bar T_{\text{cel}}$.
This is the holomorphic celestial Ward-Takahashi identity and is equivalent, via the celestial dictionary, to the subleading soft graviton theorem in four-dimensional flat spacetime. This result is consistent with \cite{Kapec:2016jld}.

We have another massless mode $\lambda=-i$ as
\begin{align}
    h_{IJ,\lambda = -i}
    &\sim V_{IJ,\Delta = 0}(p) + e^{-2 t} W_{IJ, \Delta = 2}(p).
\end{align}
From this mode, we can define an additional operator of conformal dimension $\Delta = 2$ as
\begin{align}
    T^{\prime IJ}
    \;\sim\;
    W^{IJ}_{\Delta = 2}(p)
    \;\propto\;{{\mathscr{O}^{IJ}_{\Delta = 2}}}(p)\equiv {\mathscr{T}^{\prime IJ}_{\Delta = 2}}(p).
\end{align}
The operator $\mathscr{T}^{\prime IJ}_{\Delta = 2}$ is identified with the ``dual stress tensor'' introduced in \cite{Ball:2019atb}. 
It is known in the context of celestial holography that the dual stress tensor is not, in general, a conserved current \cite{Pasterski:2021dqe}. From our de Sitter perspective, however, the reason why it does not correspond to a conserved current is not entirely transparent. A more detailed analysis is left for future work.

\section{Conclusion and Discussion}\label{Sec.6}
\subsection{Summary}
In this paper, following the same procedure as in \cite{Furugori:2025xkl}, we have established that the spin-1 gauge field and linearized graviton in $(D{+}2)$-dimensional Minkowski spacetime $\mathbb{M}_{D+2}$ correspond to spin-1 and spin-2 fields, respectively, with continuous mass spectrum on de Sitter space $\mathrm{dS}_{D+1}$ in the Bunch-Davies vacuum. 
From this setup, we were able to construct, following the same prescription as in \cite{Furugori:2025xkl}, an explicit dictionary between celestial CFT operators and cosmological operators, now extended to fields with spin. 
In particular, by focusing on the Nambu-Goldstone modes associated with the asymptotic symmetries in Minkowski spacetime, we identified the corresponding modes in the continuous spectra in de~Sitter space. 
For the Maxwell field, the corresponding modes reside in the massless sector within the continuous mass spectrum of the Proca field. For linearized gravity, viewed as a Fierz--Pauli field with a continuous mass spectrum, we found that supertranslations are encoded in the partially massless sector, whereas superrotations are realized in the strictly massless sector. These identifications allow us to define the de~Sitter boundary currents associated with these symmetries directly from the asymptotic behavior of the corresponding modes. 
We started from the Ward-Takahashi identities for these currents, and using the operator correspondence between cosmological and celestial operators, rewrote them purely in terms of celestial CFT operators. We found that the resulting identities reproduce the conformally soft theorems in celestial holography, which are known to be equivalent to the soft photon and graviton theorems in Minkowski spacetime.

\subsection{Future Directions}

\begin{enumerate}

\item \textbf{Central charge}
In this paper, we have shown that the stress tensor in de Sitter space agrees with that of the celestial CFT up to an overall constant factor. For the stress tensor in three-dimensional de Sitter space, it has been proposed that the associated central charge takes the form
\begin{align}
  c=-i\frac{3l}{2G}
\end{align}
up to an overall imaginary factor \cite{Strominger:2001pn, Balasubramanian:2002zh, Kelly:2012zc, Ouyang:2011fs}.   
On the other hand, for the celestial CFT central charge, there is a tension in the literature: in the foliation-based approach, such as in the work \cite{Pasterski:2022lsl, Ogawa:2022fhy, Capone:2024oim}, a non-vanishing central charge is suggested, whereas in other references \cite{Fotopoulos:2019vac, Banerjee:2022wht} it has been argued that the central charge vanishes. A more refined analysis of the correspondence established in this paper may help resolve this discrepancy.

\item \textbf{Source and current for dS}
In this work, we have proposed a correspondence between the stress tensor and the current in cosmological correlators and those in the celestial CFT. However, we have not provided a fully sharp prescription as to which operator, $V$ or $W$, should be identified as the current. A detailed analysis of the Minkowski boundary conditions and their precise mapping to the dS boundary conditions, leading to a justification for choosing $V$ as the current, is an important direction for future work.

\item \textbf{Subleading contribution of the fall-off}
We have shown that the partially massless large gauge transformations in de Sitter space coincide with supertranslations in Minkowski space, and we have identified the corresponding currents. We have also checked that the action of the extrapolated partially massless current on the scalar cosmological operator $W$ reproduces the weight shift of the conformal dimension of the celestial operator $\mathscr{O}$ in celestial holography.  
However, we have not explicitly computed how the same current acts on the operator $V$. Strictly speaking, this is not necessary for deriving the Ward-Takahashi identities of celestial holography, since $V$ corresponds to the shadow operator on the celestial side; once the Ward-Takahashi identities for non-shadow operators are known, those involving shadow insertions follow automatically.  
Nevertheless, from the viewpoint of consistency, it is an important open problem to explicitly determine how the current acts on $V$.

\item \textbf{Relation to cosmological soft theorem and cosmological bootstrap}
We have found that the asymptotic symmetry in Minkowski spacetime is realized as the asymptotic symmetry in de Sitter space in our mapping. The asymptotic symmetry is related to the soft theorems in Minkowski spacetime. 
It is interesting to consider its relations to the cosmological soft theorem \cite{Maldacena:2002vr}. 
In addition, it is also interesting to study the implications from consistency relations in the cosmological bootstrap (see, e.g., \cite{Arkani-Hamed:2018kmz, Baumann:2019oyu, Baumann:2020dch}) to the celestial amplitudes. 

\end{enumerate}

\section*{Acknowledgements}
We are grateful to Shinji Mukohyama, Tadashi Takayanagi, Takahiro Tanaka, and Taishi Kawamoto for useful discussions.
HF is supported by JSPS Grant-in-Aid for Scientific Research KAKENHI Grant No. JP22H01217.
NO is supported by JSPS KAKENHI Grant Number JP24KJ1372. 
TW is supported by JSPS KAKENHI Grant Number JP25KJ1621. 
The work of NO is partially supported by Grant-in-Aid for Transformative Research Areas (A) “Extreme Universe” No. 21H05187.
SS acknowledges support from JSPS KAKENHI Grant Numbers JP21K13927 and JP22H05115, and JST BOOST Program Japan Grant Number JPMJBY24E0.

\appendix
\section{Decomposition of the Gravitational Action}
\label{App.gravreduction}
We start from 
\begin{align}
  G_{IJ}dX^IdX^J
   =  e^{2\xi}\qty(d\xi^2+g_{ij}(\xi,x)\,dx^i dx^j).
\end{align}
Under the Weyl rescaling
\begin{align}
    G_{IJ} \;=\; e^{2\xi}\,\tilde{G}_{IJ},
\end{align}
the $(D{+}2)$-dimensional Ricci scalar transforms as
\begin{align}
    R'[G]
    &= e^{-2\xi}\Bigl[
        R'[\tilde{G}]
        - 2(D+1)\,\tilde{\Box}\xi
        - D(D+1)\bigl(\tilde{\nabla}\xi\bigr)^2
      \Bigr]
      \nonumber\\
    &= e^{-2\xi}\Bigl[
        R'[\tilde{G}]
        - 2(D+1)\,K
        - D(D+1)
      \Bigr],
\end{align}
where $\tilde{\Box}$ and $\tilde{\nabla}$ are computed with respect to $\tilde{G}_{IJ}$, and $K$ denotes the trace of the extrinsic curvature of the constant-$\xi$ hypersurfaces as $K_{ij}=\frac{1}{2}\partial_\xi g_{ij}$, $K=g^{ij}K_{ij}$.

The Gauss--Codazzi relation for the induced metric $g_{ij}$ on the constant-$\xi$ slices reads
\begin{align}
    R'[\tilde{G}]&=R[g] - K_{ij}K^{ij} + K^2 + 2\tilde{\nabla}_I(n^J \tilde{\nabla}_J n^I-n^I\tilde{\nabla}_Jn^J)\nonumber\\
    &= R[g] - K_{ij}K^{ij} + K^2 - 2\tilde{\nabla}_I(n^I K)
      \nonumber\\&= R[g] - K_{ij}K^{ij} + K^2 - 2\bigl(K^2 + \partial_{\xi}K\bigr),
\end{align}
where $n^I$ is the unit normal vector to the slices with respect to $\tilde{G}_{IJ}$.

Combining these expressions, we obtain
\begin{align}
    \sqrt{G}\,R'[G]
    &= e^{(D+2)\xi}\sqrt{\tilde{G}}\;
       e^{-2\xi}\Bigl[
          R'[\tilde{G}]
          - 2(D+1)\,K
          - D(D+1)
       \Bigr]
       \nonumber\\
    &= e^{D\xi}\sqrt{g}\Bigl[
          R[g]
          - K_{ij}K^{ij}
          + K^2
          - 2\bigl(K^2 + \partial_{\xi}K\bigr)
          - 2(D+1)\,K
          - D(D+1)
       \Bigr]
       \nonumber\\
    &\sim e^{D\xi}\sqrt{g}\Bigl[
          R[g]
          - K_{ij}K^{ij}
          + K^2
          - D(D-1)
       \Bigr].
\end{align}
Here $\sim$ denotes equality up to total derivative terms, i.e.\ terms that become boundary contributions after integration by parts.\footnote{
We use the following relations:
\begin{align*}
    \partial_\xi\qty(e^{D\xi}\sqrt{g})=e^{D\xi}\sqrt{g}\qty(D+K),\qquad
    \partial_\xi\qty(e^{D\xi}\sqrt{g}K)=e^{D\xi}\sqrt{g}\qty(DK+K^2+\partial_{\xi}K).\nonumber
\end{align*}
}

\section{Brief Review of Partially Massless Graviton in de Sitter}\label{FPreview}

We consider linearized gravity on a $(D+1)$-dimensional maximally symmetric spacetime \cite{Aragone:1971kh, Aragone:1979bm, Kan:2022fcj}. The action for the metric fluctuation $h_{ij}$ about the background metric $g_{ij}$ is
\begin{align}
  S
  = \int [d^{D+1}x]\qty[
    -\frac{1}{2}\nabla_k h_{ij}\nabla^k h^{ij}
    + \nabla_i h_{jk}\nabla^j h^{ik}
    - \nabla_i h^{ij}\nabla_j h
    + \frac{1}{2}\nabla_k h\nabla^k h
    + \frac{R}{D+1}\qty(h_{ij}h^{ij} - \frac12 h^2)
  ],
\end{align}
where $\nabla$ is the Levi-Civita connection of $g_{ij}$ and $h \equiv g^{ij}h_{ij}$.
This action is invariant under the linearized diffeomorphisms
\begin{align}
  \delta h_{ij} = \nabla_i \xi_j + \nabla_j \xi_i.
\end{align}

This symmetry is broken by the Fierz--Pauli mass term
\begin{align}
  \mathcal{L}_{\text{mass}}
  = \frac{1}{2}m^2\qty(h^{ij}h_{ij} - h^2),
\end{align}
with nonzero Fierz--Pauli mass $m^2\neq 0$.
In this case, we have the (linearized) Bianchi identity
\begin{align}
  \nabla^{i} h_{ij} = \nabla_j h,
\end{align}
and the following constraint,\footnote{We set the de Sitter radius to unity.}
\begin{align}
  \qty[m^2 - (D-1)]\,h = 0.
\end{align}
Therefore, for
\begin{align}
  m^2 \neq D-1,
\end{align}
the trace vanishes.

At the special mass value\footnote{In de Sitter space, this Fierz--Pauli mass saturates the lower end of the Higuchi bound~\cite{Higuchi:1986py},
\begin{align}
  m^2 \geq D-1.
\end{align}} corresponding to a partially massless spin-2 field \cite{Deser:1983tm, Deser:1983mm, Deser:2001us,Deser:2001wx},
\begin{align}
  m^2 = D-1,
\end{align}
an additional scalar gauge invariance appears,
\begin{align}
  \delta_f h_{ij}
  = \qty(\nabla_i\nabla_j + g_{ij})f.
\end{align}

The corresponding conformal dimensions of the partially massless graviton and its shadow operator are
\begin{align}
  \Delta_+ = D-1,\qquad\Delta_-=1.
\end{align}
Thus, in global coordinates $(t,x^\alpha)$, we obtain the following late-time falloff behavior:
\begin{align}
  h_{\alpha\beta}(t,x)
  \sim e^{-t} W_{\alpha\beta}(x) + e^{-(D-1)t} V_{\alpha\beta}(x).
\end{align}
In particular, we regard $V_{\alpha\beta}$ as a dual current in the main text, and we have a partial conservation law \cite{Dolan:2001ih}
\begin{align}
  \nabla_\alpha \nabla_\beta V^{\alpha\beta} = 0,
\end{align}
where $\nabla_\alpha$ is the covariant derivative with respect to the metric on sphere $S^D$.
Including the contributions of other local operators, this current generates a transformation according to
\begin{align}
  \int d^Dx\, f(x)\,
  \ev{\nabla_\alpha \nabla_\beta V^{\alpha\beta}(x)\prod_n O_n(x_n)}
  = \ev{\delta_f O_m(x_m)\prod_{n\neq m}O_n(x_n)}.
\end{align}

\bibliographystyle{utphys} 
\bibliography{ref} 

\end{document}